\documentclass[aps,prl,twocolumn,superscriptaddress]{revtex4}
\usepackage{graphicx}
\usepackage{dcolumn}
\usepackage{bm}
\usepackage{amsmath}
\usepackage{multirow}
\usepackage{upgreek}

\begin{document}

\title {Evidence of Lifshitz transition in thermoelectric power of ultrahigh mobility bilayer graphene}

\author{Aditya~Jayaraman}
\email{jaditya@iisc.ac.in}
\affiliation{Department of Physics, Indian Institute of Science, Bangalore 560 012, India.}
\author{Kimberly~Hsieh}
\affiliation{Department of Physics, Indian Institute of Science, Bangalore 560 012, India.}
\author{Bhaskar~Ghawri}
\affiliation{Department of Physics, Indian Institute of Science, Bangalore 560 012, India.}
\author{Phanibhusan S. Mahapatra}
\affiliation{Department of Physics, Indian Institute of Science, Bangalore 560 012, India.}
\author{Arindam~Ghosh}
\email{arindam@iisc.ac.in}
\affiliation{Department of Physics, Indian Institute of Science, Bangalore 560 012, India.}
\affiliation{Centre for Nano Science and Engineering, Indian Institute of Science, Bangalore 560 012, India.}

\begin{abstract}


\textbf{Resolving low-energy features in the density of states (DOS) holds the key to understanding wide variety of rich novel phenomena in graphene based 2D heterostructures. Lifshitz transition in bilayer 	
graphene (BLG) arising from trigonal warping has been established theoretically and experimentally. Nevertheless, the experimental realization of its effects on the transport properties has been
challenging because of its relatively low energy scale ($\sim 1$~meV). In this work, we demonstrate that the thermoelectric power (TEP) can be used as an effective probe to investigate fine
changes in the DOS of BLG. We observe additional entropy features in the vicinity of the charge neutrality point (CNP) in gapped BLG. This apparent violation of Mott 
formula can be explained quantitatively by considering the effects of trigonal warping, thereby serving as a possible evidence of a Lifshitz transition.}

\end{abstract}
\maketitle

	Lifshitz transition (LT) involves changes in the topology of Fermi surface that result in anomalies in the electronic properties of a wide variety of systems ~\cite{lifshitz1960}. LT is generally accompanied by the divergence of DOS resulting in van-Hove singularities (vHS). In many cases therefore, LT is associated with the onset of correlated electronic phases such as superconductivity~\cite{SC_cuprates,FeSe_SC}, magnetism~\cite{LT_magnet1,LT_magnet2,LT_magnet3} and plays a vital role in the properties of Weyl semi-metals~\cite{weylnbas,weylmote2}. In addition to the broken symmetry states, LT also leads to the modification of the transport properties at the transition~\cite{LT_resistivity,snseTEP}. Bilayer graphene (BLG) has a low-energy parabolic bandstructure with zero bandgap, while the application of a transverse electric field ($D$) opens up a bandgap resulting in a `mexican hat' dispersion. BLG provides a tunable test case for LT, where the addition of skew inter-layer hopping ($\gamma_{3}$) term (Fig.~1a) in the tight-binding Hamiltonian trigonally distorts its low-energy bandstructure ~\cite{mccann2007low,mccannkoshino}. The parabolic bandstructure with a single Fermi surface transforms into four Fermi surface pockets in the vicinity of K and K' points (Fig.~1b). They coalesce to form a single Fermi surface resulting in a LT at an energy $E_{\mathrm{L}}$ (Fig.~1c).

	Previous reports on the observation of LT in the transport properties of BLG have been scarce because of its relatively low energy scale ($\sim 1$~meV), rendering these effects negligible at higher temperatures~($>10$~K)~\cite{science2011reconst,varlet2014,oka2019ballistic}. Furthermore, the density required to shift the chemical potential to $E_{\mathrm{L}}$ is very small ($n_{\mathrm{L}} \sim 5 \times~10^{10}$~cm$^{-2}$). Hence, experimental demonstration of such a transition is limited by fluctuation- and inhomogeneity-induced band smearing due to the presence of disorder. Therefore it is imperative to enhance the strength of trigonal warping in order to make the experimental observation of LT feasible. Recent reports indicate that the strength of trigonal warping can be enhanced by the application of transverse electric field or presence of tensile or shear strain ~\cite{mucha2011strained,varlet2014}. 
In addition to the enhancement of $E_{\mathrm{L}}$, there is also a need for a measurement technique which is sensitive to the variation of DOS.

 \begin{figure} 
\includegraphics[width=1.0\linewidth]{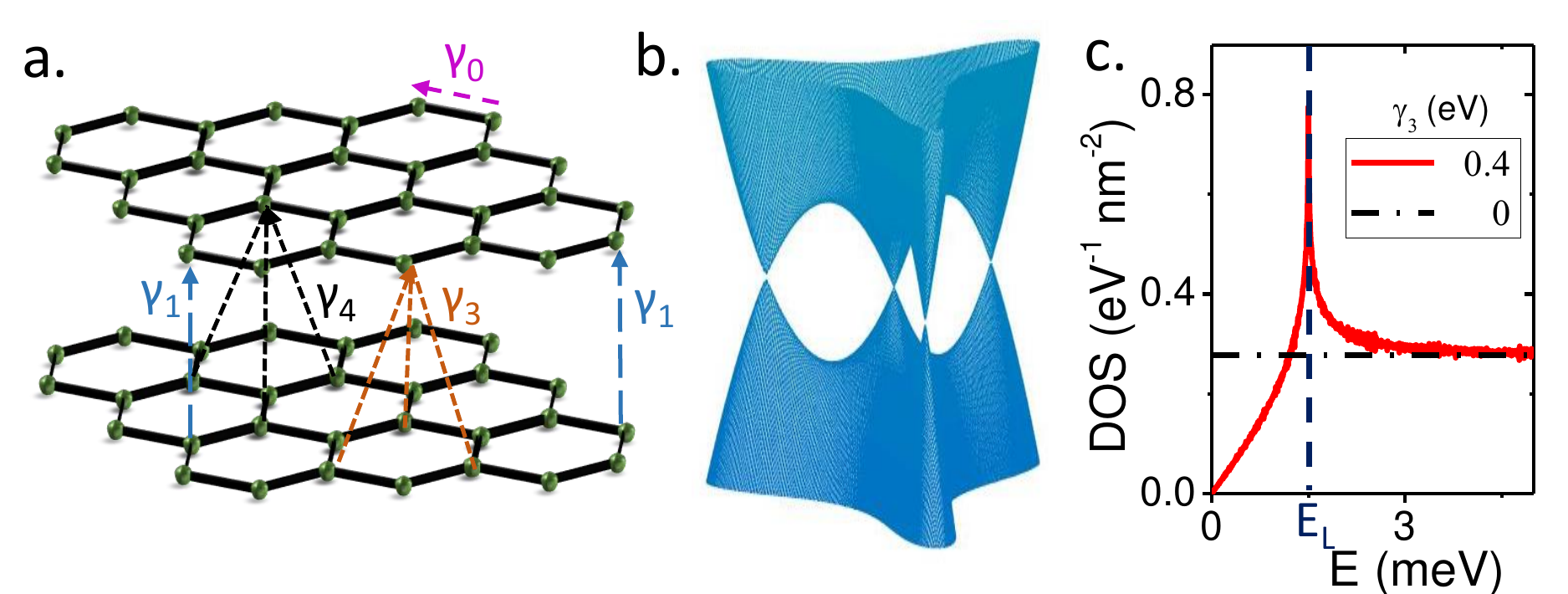}
\caption{\textbf{Lifshitz transition in bilayer graphene:} \textbf{a.}~Schematic of lattice structure of Bernal stacked BLG. The different electron hopping terms are shown. \textbf{b.} Bandstructure of BLG showing trigonal distortion of Fermi surface. \textbf{c.} DOS of BLG with the inclusion of $\gamma_{3}$ showing Lifshitz transition at energy $E_{\mathrm{L}}$. Black dashed line represents the constant DOS in the absence of $\gamma_{3}$ in the Hamiltonian.}
\end{figure}
\label{Fig0}

Here, we observe unambiguous signature of LT in BLG via the measurement of thermoelectric power (TEP). We demonstrate the extreme sensitivity of TEP to the band structure, and that it can provide information about the electronic structure which cannot be determined by conductance measurements alone~\cite{TEPnanost}. 
Earlier studies on the TEP of BLG have shown that it follows the well-understood semi-classical Mott relation for thermopower ~\cite{philipkim2009TE,wang2011enhanced,phani2017}:
\begin{equation}
S_{\mathrm{Mott}}(\mu,T) = \frac{\Delta V}{\Delta T} = \frac{\pi^{2}k_{\mathrm{B}}^{2}T}{3e} \biggl (\frac{d(ln(G))}{dE}\biggr )_{E=\mu}
\end{equation}
where $\Delta V$ is thermoelectric voltage, $G$ is the energy-dependent conductance of the sample, $T$ is the sample temperature and $\Delta T$ is the temperature gradient.   
From the Mott relation, it is evident that small changes in conductance leads to large changes in the TEP since it is proportional to the derivative of DOS within the semi-classical Boltzmann theory. Our experiments reveal new features observed in TEP which can be attributed to trigonal warping, enhanced due to the application of transverse electric field and inbuilt sample strain, thereby confirming TEP as an ideal transport coefficient to resolve low-energy features in DOS.

\begin{figure} 
\includegraphics[width=01\linewidth]{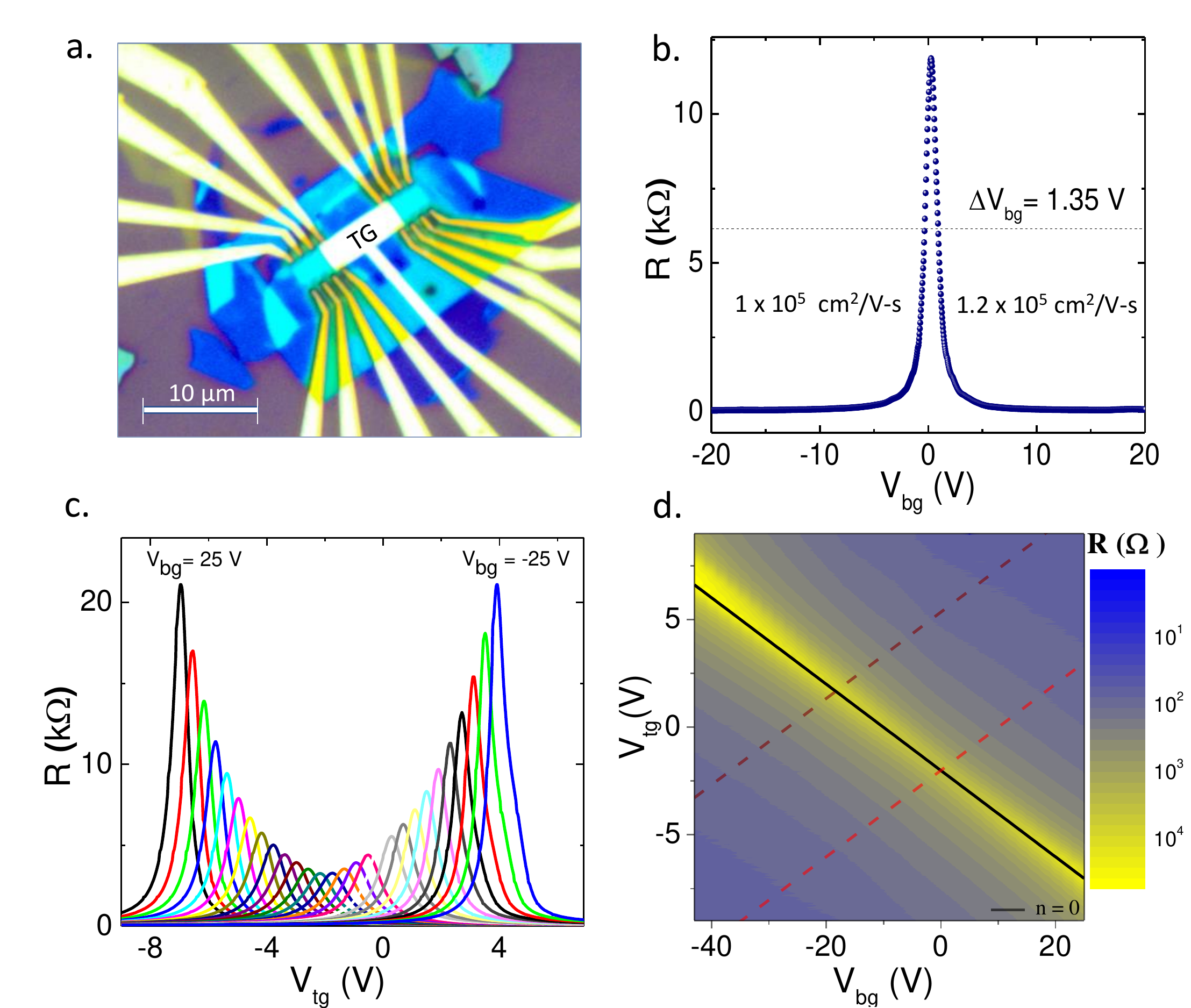}
\caption{\textbf{Device structure and characterization:} \textbf{a.}~The optical micrograph of hBN encapsulated BLG FET device with edge contacts. \textbf{b.} $ R-V_{bg}$ at 8~K. The field effect mobility of the sample is $1 \times 10^{5}$ and $1.2 \times 10^{5}$~cm$^{2}$/V-s for hole and electron sides respectively. FWHM of $R-V_{g}$ is 1.35~V corresponding to the carrier homogeneity of $9 \times 10^{10}$~cm$^{-2}$. \textbf{c.} Transfer characteristics of the device at different values of $V_{bg}$ from -25~V to 25~V in steps of 2~V ($T=77$ K) \textbf{d.} Colour plot of resistance in the ($V_{bg},V_{tg}$) phase space. }
\end{figure}
\label{Fig1}

	 The boron-nitride (hBN) encapsulated BLG device was fabricated in a dual-gated geometry in order to have independent control over the band-gap and number density. Fig.~2a shows the schematic of a BLG dual-gated field effect transistor (FET) device. Here, hBN is used as the dielectric for the top gate above BLG channel. In order to study the intrinsic channel characteristics with minimal effects from electrical contacts, the probes have been deposited by edge contacting method as described in~\cite{edge}. The local top gate (Fig.~2a) covers the main channel ($6.5~\upmu$m) whereas the channels outside the main channel are gated by the global SiO$_{2}$/Si$^{++}$ back gate alone. The graphene channel outside the top-gated region is used as a heater to set up a temperature gradient across the main channel for thermoelectric measurements (see Fig.~3a).   
        Fig.~2b represents the back gate transfer characteristics of the device. The field effect mobilities of the channel $\mu_{e}\approx1.2 \times 10^{5}$~cm$^{2}$/Vs at 8~K. The width of the resistance versus gate voltage peak at the CNP gives an estimate of the charge inhomogeneity resulting from electron-hole puddle formation at low densities~\cite{dean2010hbn}. In Fig.~2b, the full-width at half-maximum (FWHM) of R($V_{bg}$) is $\sim$ 1.35~V, giving an upper bound for disorder-induced carrier density fluctuation of $\delta n \lesssim 9 \times 10^{10}~$cm$^{-2}$. The higher mobility of the sample and the smaller value of $\delta n$ signify the minimal presence of external disorder and impuritires.  Fig.~2c shows the transfer characteristics of the dual-gated BLG device. The top gate was swept by keeping the back gate fixed at different values between -25~V and 25~V. The increase in the resistance at the CNP with varying back gate voltage confirms the increase in band gap. The phase-space colour plot of the same data is plotted in Fig.~2d. The solid black diagonal line corresponds to positions of CNP at various values of electric field $D$ ($n =0$), while the dashed lines correspond to the locus of $D= 0$~V/nm (red) and $D = -0.1$~V/nm (brown).

	TEP is obtained from the ratio of thermoelectric voltage and the temperature gradient ($V_{2\omega}/\Delta T$) across the channel. The measurement schematic for TEP is presented in Fig.~3a. A temperature gradient is generated across the channel via Joule heating by passing a sinusoidal current ($I_{\omega}$) along a channel outside the top gated region. The resulting second harmonic thermoelectric voltage ($V_{2\omega}$) is measured across the top gated channel at different transverse electric fields ($D$) with varying number density. 
Since the temperature gradient set up is proportional to Joule heating ($\propto I^{2}$), $V_{2\omega}$ also scales with $I_{\omega}^{2}$ for the range of heating currents used ($100-600$~nA), ensuring that the TEP measurements are performed in linear response regime ($\Delta T\ll T$). This is confirmed in Fig.~3b and its inset. 

\begin{figure*} 
\includegraphics[width=0.8\linewidth]{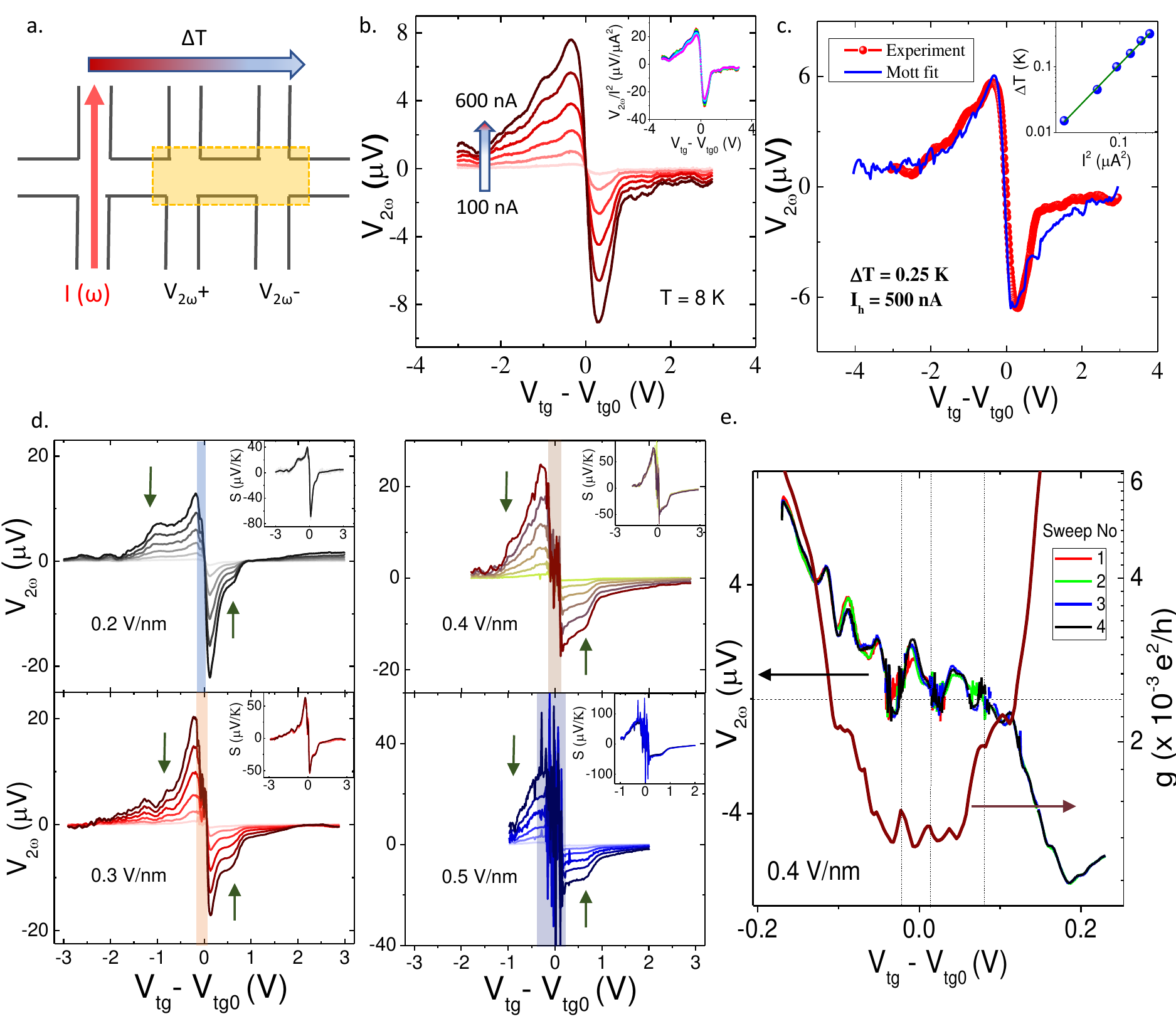}
\caption{\textbf{Thermoelectric measurements:}  \textbf{a.} Contact configuration for thermopower measurements. \textbf{b.} $V_{2\omega}$ as a function of top gate voltage for different values of heating currents at 8~K and $D= 0$~V/nm. Inset shows the normalized $V_{2\omega}/I^{2}$ vs $V_{g}$. \textbf{c.} Mott fitting of  $V_{2\omega}$ ($I_{h} =$ 500~nA) with $\Delta T$ as a fitting parameter. The extracted values of  $\Delta T$ at various heating currents is presented in the inset.  \textbf{d.} Gate voltage dependence of thermovoltage at different transverse electric field ($T=8$~K) for various heating currents. The shaded portion indicates the region of fluctuating thermopower. Insets show the thermopower calculated for non-zero $D$ with the same calibration of $\Delta T$ estimated for $D=0$ for different heating currents. \textbf{e.}  Magnified version of thermoelectric voltage near the CNP for $D=0.4$~V/nm showing repeatability of fluctuations (shaded box in Fig.~3d). The corresponding fluctuations in conductance are also shown.}
\end{figure*}
\label{Fig2}

Since the thermopower measurements are done at low temperatures (8~K), graphene resistance thermometry is not a useful method to estimate $\Delta T$ because of the weak dependence of graphene resistance on temperature~\cite{taychatanapatgappedblg}. Hence, we estimate $\Delta T$ by fitting the Mott formula to the measured thermovoltage $V_{2\omega}$ at $D=0$. 
Fig.~3c shows the Mott fitting for $V_{2\omega}$ ($D=0$~V/nm and $I_{h}=$~500~nA) with $\Delta T$ as a fitting parameter. The fit gives $\Delta T \approx0.25$~K. Similar Mott fits were obtained for different values of $I_{h}$ and the extracted values of $\Delta T$ are plotted vs $I_{h}^{2}$ in the inset of Fig.~3c. The linearity of the estimated values of $\Delta T$ confirms the self-consistency of this method. This method of determining $\Delta T$ is justified under the assumption that the system obeys Mott relation, which is indeed true for BLG~\cite{wang2011enhanced,nam2010blg}. Same calibration of $\Delta T$ was subsequently used for thermopower estimation at non-zero $D$ upon scaling with the corresponding heater resistance. The calculated $S$ for different heating currents (with same calibration of $\Delta T$) collapse on top of one another, providing additional support to the method (insets in Fig.~3d). 
\begin{figure*}
\includegraphics[width=1\linewidth]{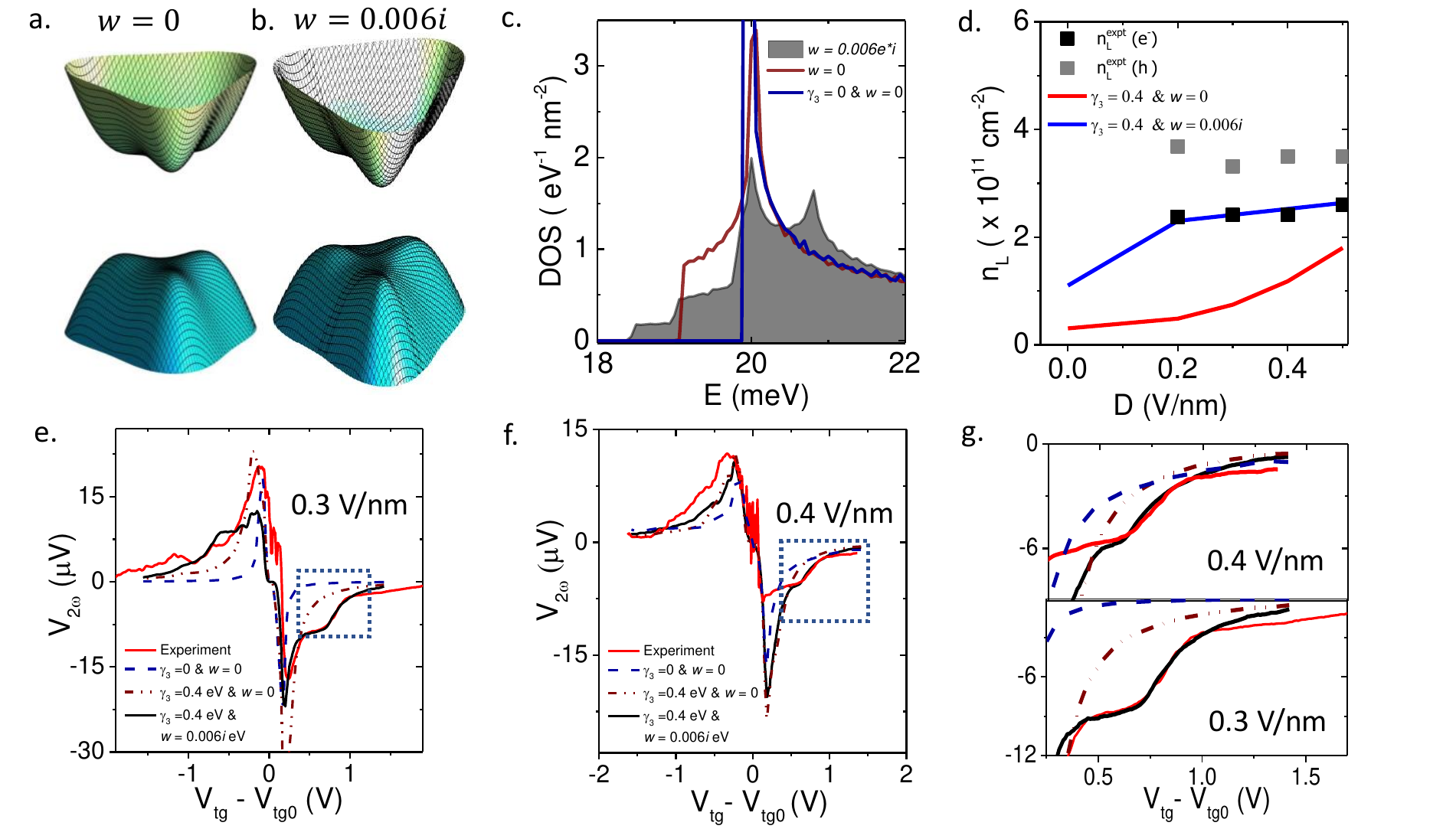}
\caption{\textbf{Mott fitting of TEP:} Bandstructure of BLG with $\gamma_{3}$ = 0.4~eV, $D = 0.4$~V/nm  \textbf{a.} $w = 0$~meV and \textbf{b.} $w = 6i$~meV. \textbf{c.} DOS of BLG ($D = 0.4$~V/nm). Shaded region shows DOS with the inclusion of strain. \textbf{d.} Calculated $n_{\mathrm{L}}$ vs $D$ compared with the experimental data. Mott fitting of TEP with various approximations of DOS for \textbf{e.} $D = 0.3$~V/nm and \textbf{f.} $D = 0.4$~V/nm. \textbf{g.} Magnified version of (e) and (f) to show fitting of Mott formula in the electron doped region.}
\end{figure*}
\label{Fig3}

Fig.~3d shows the evolution of thermoelectric voltage with increasing gap for different values of heating current ($100-600$~nA). Previous reports on gapped BLG~\cite{wang2011enhanced,TEPionicgating} show considerable enhancement in thermopower as a function of $D$, which is consistent with our observation (see insets of Fig. 3d and Fig.~S1). This enhancement is due to the change in band curvature at band edges with increasing $D$~\cite{wang2011enhanced}. Moreover, increasing $D$ introduces sharp fluctuations in thermopower when the chemical potential is within the gap. These fluctuations occur very close to $n = 0$, and the gate voltage window, over which the fluctuations occur, increases with the opening of band gap. Upon closer inspection, these fluctuations seem to be reproducible with multiple sweeps overlapping on top of each other (Fig. 3e). Universal conductance fluctuations can be ruled out because the channel is strongly localized in this gate voltage window (G $\ll e^{2}/h$). It is well known that the transport in strongly gapped regime of BLG occurs via hopping across the localized trap states within the gap at low temperatures~\cite{taychatanapatgappedblg}. Fluctuations in conductance in the strongly localised regime have been studied extensively and are known to occur when the transport occurs via hopping mechanism~\cite{aamir2018}. The hopping probability between two sites is sensitive to the Fermi energy and the traps are exponentially activated. Therefore with varying Fermi energy, a different pair of hopping sites may be activated leading to large but repeatable fluctuations in the resistance, which are magnified in the TEP measurements.

The new and most striking observation in our measurements is the presence of thermopower `plateaus' in the vicinity of CNP (marked by arrows in Fig. 3d). These highly reproducible features, present on either side of CNP, are reflected only in the TEP with no corresponding features in resistance measurements (see Fig.~S2). The plateaus, though not visible for $D=0$~V/nm, are observed for all measured values of $D>0$ and the density at which the plateaus occur is weakly dependent on $D$. They occur at $n_{\mathrm{L}}\approx~2.5\times 10^{11}$~cm$^{-2}$  and  $n_{\mathrm{L}}\approx~-3.3\times~10^{11}$~cm$^{-2}$ on the electron and hole side respectively for all values of $D$.  
These plateaus are reminiscent of `entropy spikes' expected to occur at Lifshitz transition in 2D systems like germanene and silicene~\cite{galperin2018entropy,germanine2018}. Recent theoretical studies have also predicted the occurrence of similar secondary maxima in thermopower as a function of density in ballistic BLG system because of LT~\cite{suszalski2018lifshitz,suszalski2019TEP}. These features cannot be explained by DOS without the inclusion of $\gamma_{3}$, thereby giving an impression of an apparent violation of Mott law.   
        
In order to interpret the presence of the plateaus in TEP measurements at characteristic number density, tight-binding calculations of BLG with the inclusion of trigonal warping were performed. The four-band effective Hamiltonian for BLG for low energy excitations is given by~\cite{mccannkoshino}:

\begin{equation}
\mathbf{\hat{H}} =
\begin{bmatrix}
\frac{\xi}{2}U &  v_{3}\mathbf{\hat{\pi}} +w & 0 & v_{f}\mathbf{\hat{\pi^{\dagger}}}  \\
v_{3}\mathbf{\hat{\pi}^{\dagger}} +w^{\ast} & -\frac{\xi}{2}U & v_{f}\mathbf{\hat{\pi}} & 0\\
0 &  v_{f}\mathbf{\hat{\pi^{\dagger}}} & -\frac{\xi}{2}U & \xi\gamma_{1} \\
 v_{f}\mathbf{\hat{\pi}} & 0 & \xi\gamma_{1} & \frac{\xi}{2}U
\end{bmatrix}
\end{equation}
where, $\xi=1$ ($\xi=-1$) refers to K(K') valley, $\gamma_{1}$ (=0.381~eV) is the interlayer coupling, $\gamma_{3}$ (=0.4~eV) is the skew interlayer coupling, $U$ is the inter-layer asymmetry term and $\hat{\pi} = p_{x} + ip_{y}$. Here, $v_{3}$ (=$\frac{\sqrt{3}a}{2\hbar}e\gamma_{3}$) where $a$ is the lattice constant of BLG. $w$ is a complex gauge parameter that represents the magnitude of homogeneous strain on BLG~\cite{mucha2011strained}. 

The addition of skew interlayer hopping term modifies the low energy bandstructure of gapped BLG as well. Unlike the fragmentation of Fermi surface into four pockets as in the case of BLG at $D=0$~V/nm, the conduction and valence bands are trigonally distorted near the band edge with three disconnected Fermi pockets as shown in Fig.~4a for $D=0.4$~V/nm. Therefore, the interplay of interlayer asymmetry and trigonal warping produces LT close to the band edge, as shown in the DOS plot (Fig.~4c (brown)). $n_{\mathrm{L}}$ is calculated by integrating the DOS over the energy range from $0$ to $E_{\mathrm{L}}$. 
 Fig.~4d shows the calculated values of $n_{\mathrm{L}}$ as compared to the experimentally observed values. Clearly, calculated values of $n_{\mathrm{L}}$ (red trace) does not match the experimental values and show a monotonous increase with $D$ in contrast to a relatively invariant value obtained in the experiment. Therefore, a simple addition of trigonal warping term fails to explain the weak dependence of $n_{\mathrm{L}}$ with $D$. 

Multiple theoretical reports predict the enhancement of $E_{\mathrm{L}}$ with strain~\cite{mucha2011strained,varlet2015strain}. Strain can be induced either intentionally or unintentionally due to the presence of mechanical deformation of the sample, either by uniaxial strain or by interlayer shear shift caused during the sample preparation~\cite{varlet2015strain,TEPstacking2019,2011strainedBLG}. Graphene-hBN van der Waals heterostructures are known to host strain of $\sim$~0.6-1\%~\cite{pan2012biaxial,vincent2018nanostrain} because of thermal cycling, local corrugations like bubbles, or due to lattice commensuration~\cite{woods2014commensurate}, which may lead to the formation of domain boundary solitons. In addition, epitaxially grown graphene-hBN heterostructures can host strain upto~1.8\% because of lattice matching with hBN~\cite{davies2018epitaxystrain}.
Theoretically, strain can be introduced into the bandstructure by incorporating a complex parameter $w$ in the Hamiltonian (Eq. 2). It has been estimated that $1\%$ strain in monolayer graphene results in $|w| \sim 6$~meV ~\cite{mucha2011strained}. Fig. 4b represents the bandstructure of BLG with the inclusion of strain ($w = 6i$~meV). Strain induced deformation of bandstructure gives rise to distinctive features in DOS. Addition of strain breaks the symmetry between the side cones resulting in multiple saddle points and parabolic minima, which give rise to multiple Lifshitz transitions as shown Fig.~4c (shaded region). It has to be noted that the strain induced spectral changes are very sensitive to the parameter $w$ (see supplementary information). 
The calculated values of $n_{\mathrm{L}}$ obtained upon the inclusion of strain ($w = 6i$~meV) are nearly invariant for $D > 0$ (blue trace in Fig.~4c) and match well with the experimental values in the electron side.  

Fig. 4e and 4f show the fitting of Mott formula (Eq.~1) of the data calculated with the inclusion of both $\gamma_{3}$ and $w$ in DOS for $D=0.3$~V/nm and $D=0.4$~V/nm. It shows excellent fit and reproduces the plateau in the electron doped region (Fig. 4g). The Mott fits calculated without these modifications in DOS are represented with dashed lines and they fail to reproduce the plateaus. Although the addition of strain replicates the plateaus in the hole doped region as well, the calculated value of $n_{\mathrm{L}} $ is electron-hole symmetric and hence does not match the experimental value for $n < 0$, possibly because of inhomogeneneity in the strain which is not considered in our calculations.  
 Similar fitting of the Mott relation have been performed for other values of $D$ (Fig.~S4). Although addition of strain in the form of a gauge field ($w_{3}$) quantitatively describes the plateaus, correlation effects have been neglected in our calculation and cannot be discounted. Coulomb interaction is predicted to further enhance the strain-induced effects, leading to a larger value of $w$ even for negligible values of strain ~\cite{mucha2011strained}. The interaction driven reconstruction of low energy bandstructure could also explain the larger Lifshitz energy $E_{\mathrm{L}}$~\cite{science2011reconst}.
\begin{figure} 
\includegraphics[width=1\linewidth]{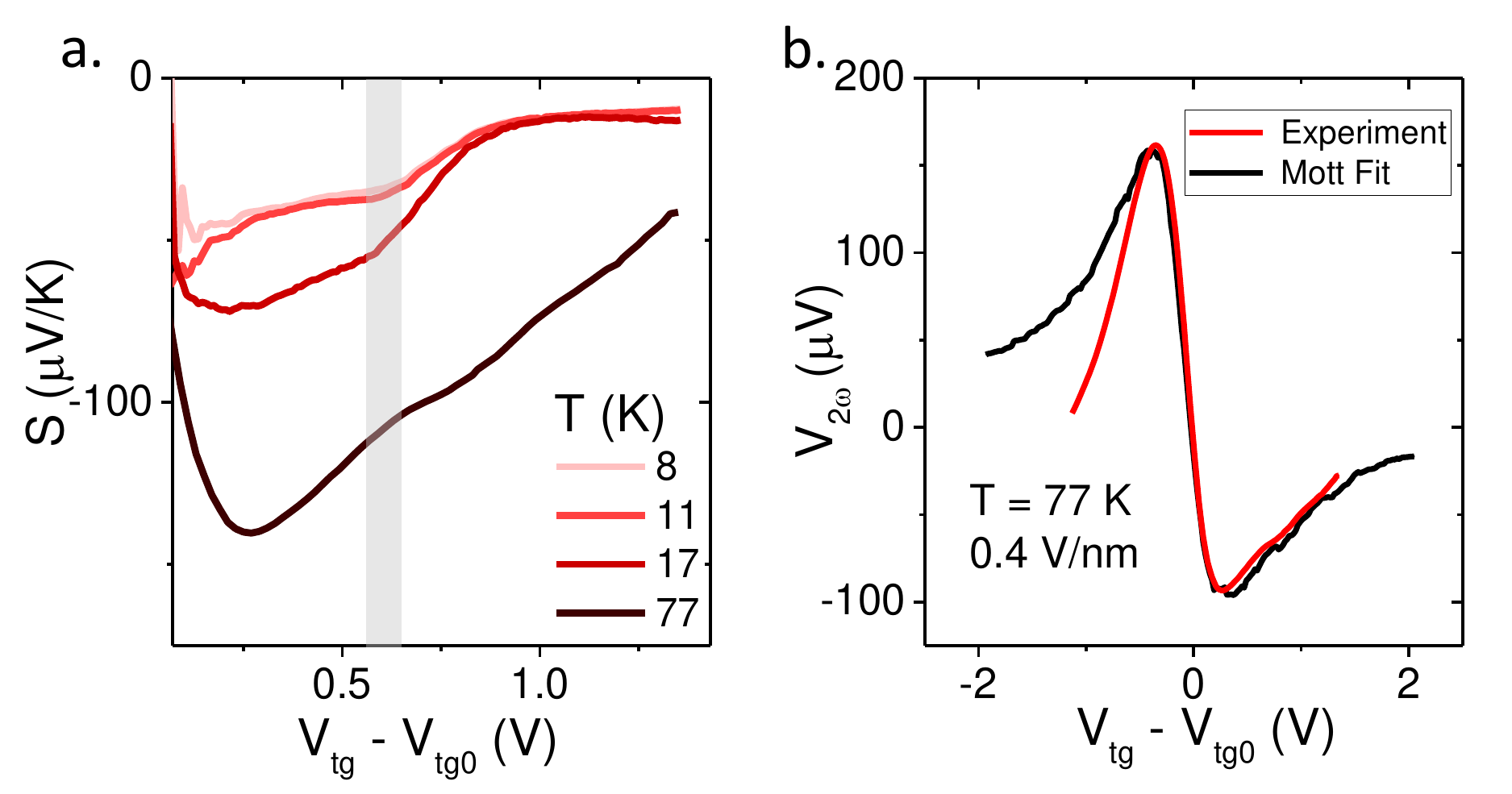}
\caption{\textbf{Temperature dependence:} \textbf{a.} Plot of  $V_{2\omega}$ vs $V_{tg}$ at various temperatures for $D=0.4$ V/nm. \textbf{b.} Mott fitting of TEP for $D = 0.4$ V/nm at 77 K ($\gamma_{3} = 0$ and $w = 0$).}
\end{figure}
\label{Fig4}

Finally, Fig.~5a shows the temperature dependence of TEP at $D=0.4$~V/nm. While the plateaus are smoothened over and are not observed beyond 17~K, there is no evidence of these features at 77~K because of temperature-induced broadening of DOS. The absence of plateaus in the case of $D = 0$~V/nm can also be explained due to the lower value of $n_{\mathrm{L}}$ (Fig.~4c)  resulting in the smearing of the plateau even at 8~K.
 There is no evidence of strain relaxation effects with increasing temperature since the plateaus occur at approximately same number density for all values of $T$ (shaded region in Fig.~5a). Fig. 5b shows Mott fitting of thermoelectric voltage at $D=0.4$~V/nm at 77~K which shows a reasonably good fit even without the addition of $\gamma_{3}$ or strain, indicating the inability to resolve the low energy features at higher temperatures.
In conclusion, we demonstrate the use of thermoelectric measurements as an alternative probe to investigate the fine details of bandstructure in BLG. We observe reproducible thermopower fluctuations within the bandgap of BLG. 
In addition, the TEP in high-mobility device enables us to detect the possible evidence of Lifshitz transition enhanced by the presence of strain as well as bandgap, which is not easily captured in conductance measurements. 

The authors thank DST for financial support.

\end{document}


\title{Supplementary Information: Evidence of Lifshitz transition in thermoelectric power of ultrahigh mobility bilayer graphene}

\author{Aditya Jayaraman}
\email{jaditya@iisc.ac.in}
\affiliation{Department of Physics, Indian Institute of Science, Bangalore 560 012, India.}
\author{Kimberly Hsieh}
\affiliation{Department of Physics, Indian Institute of Science, Bangalore 560 012, India.}
\author{Bhaskar~Ghawri}
\affiliation{Department of Physics, Indian Institute of Science, Bangalore 560 012, India.}
\author{Phanibusan~S.~Mahapatra}
\affiliation{Department of Physics, Indian Institute of Science, Bangalore 560 012, India.}
\author{Arindam~Ghosh}
\email{arindam@iisc.ac.in}
\affiliation{Department of Physics, Indian Institute of Science, Bangalore 560 012, India.}
\affiliation{Centre for Nano Science and Engineering, Indian Institute of Science, Bangalore 560 012, India.}

\maketitle

\renewcommand{\thefigure}{S\arabic{figure}}

\subsection{\large{\textbf{1.} Thermopower as a function of transverse electric field ($D$)}}

Fig.~S1(a) shows thermoelectric power (TEP) as a function of $V_{tg}$ for different values of $D$ at 8 K. Increase in $D$ results in an increase in DOS near the band edges because of the opening of band gap. Therefore, the enhancement of TEP as a function of $D$ is associated with sharp changes in the band curvature at the band edges~\cite{wang2011enhanced}. 
\begin{figure}[H]
\includegraphics[width=1\linewidth]{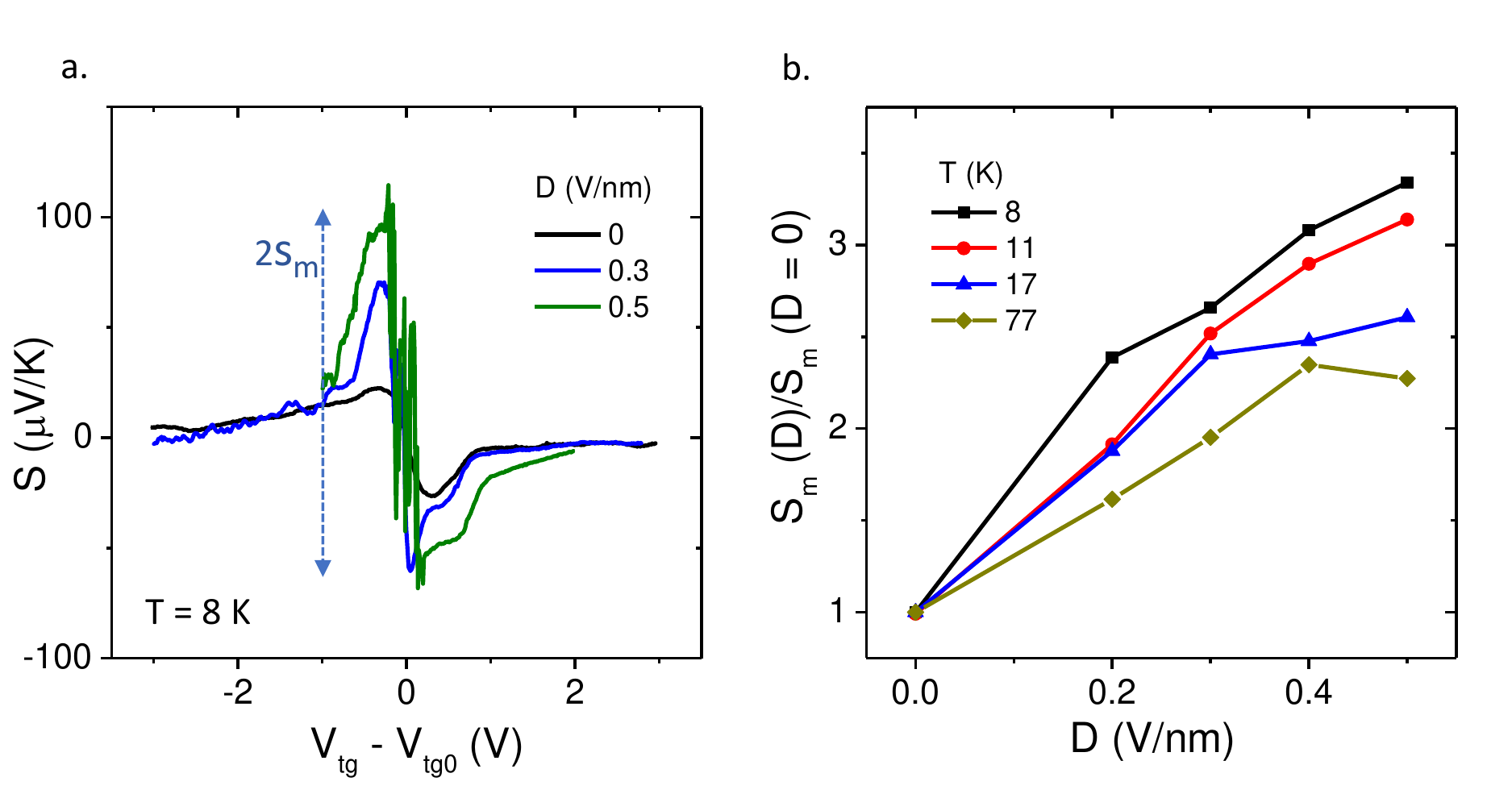}
\caption{\textbf{a.} Thermopower as a function of gate voltage at different values of $D$ ($T=8$ K). The total magnitude of TEP is defined as $2S_{m}$ (marked with arrow for 0.5 V/nm).  \textbf{b.} $S_{m}(D)/S_{m}(0)$ vs $D$ for four different temperatures showing TEP enhancement by a factor of $\sim3$. }
\end{figure}
\label{FigS1}

Fig.~S1(b) shows the plot of $\frac{S_{m}(D)}{S_{m}(0)}$ as a function of $D$. There is an enhancement of TEP by a factor of three upon increasing $D$ at 8 K (Fig. S1(b)). The enhancement factor reduces to $\sim2$ (at 77 K) with the increase in temperature. Temperature-induced broadening reduces the relative increase of the factor $\frac{1}{R} \frac{dR}{dV_{g}}$ in the Mott formula thereby resulting in overall reduction in the TEP enhancement~\cite{wang2011enhanced}.
\newpage
\subsection{\large{\textbf{2.} Comparison of TEP with conductance measurements}}

Fig.~S2 shows the TEP measurements at various values of $D$ along with resistance measurements. While TEP plateaus appear at characteristic number densities on either side of CNP, no corresponding features were observed in $R-V_{g}$ confirming TEP as a sensitive probe of DOS in comparison with conductance measurements. 

\begin{figure}[H]
\includegraphics[width=1\linewidth]{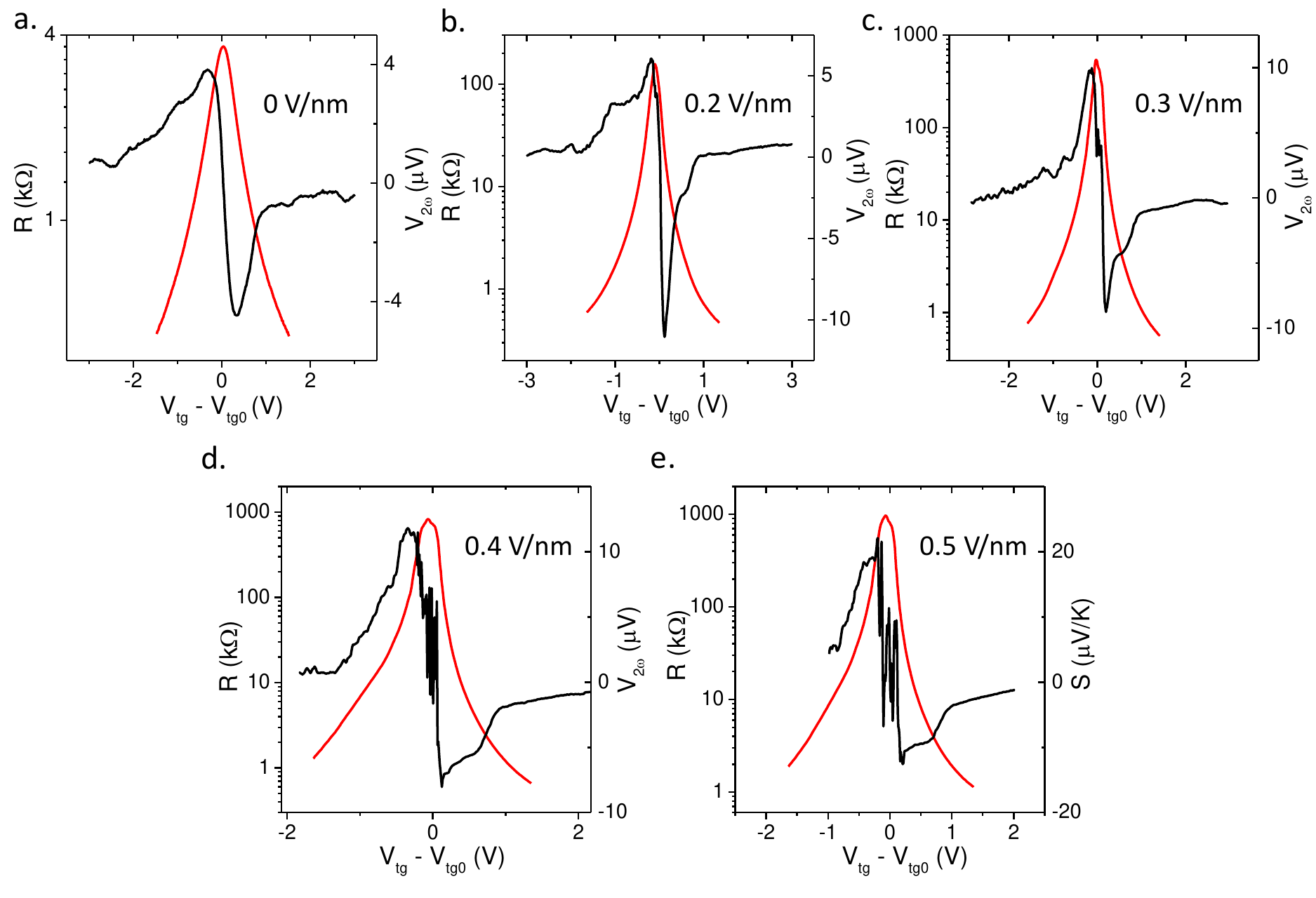}
\caption{ TEP vs $V_{tg}$ at various values of $D$ in comparison with resistance measurements.}
\end{figure}
\label{FigS2}

\subsection{\large{\textbf{3.} Sensitivity of density of states (DOS) to the complex strain parameter ($w$)}}

The bandstructure and the DOS of BLG is very sensitive to the gauge parameter $w$ used in Eq. 2 (main text). $w$ is a complex parameter used to quantify the magnitude of strain in BLG~\cite{mucha2011strained}. $w$ is given by

\begin{equation}
w = \frac{3}{4} (\eta_{3} - \eta_{0})\gamma_{3}e^{2\theta i}(\delta - \delta^{'}) - \frac{3}{2}\gamma_{3}\eta_{3}e^{i\phi}\frac{\delta r}{r_{AB}}
\end{equation}  

where,
$\delta r=(\delta x,\delta y)$ is the shift in the top layer with respect to the bottom layer due to the shear deformation,
$\tan(\phi) = \frac{\delta y}{\delta x}$,
$\eta_{0,3} = \frac{r_{AB}}{\gamma_{0,3}}\frac{\partial{r_{AB}}}{\partial{\gamma_{0,3}}}$ is the change in interlayer and skew interlayer hopping terms with respect to change in the coupled carbon atom distances,
$\delta$ and $\delta^{'}$ are the eigen values of strain tensor when a strain is applied with the principal axis at an angle $\theta$ from the coordinate axis (Fig~S3(a))~\cite{varlet2015strain}.
\newpage
Therefore $w$ is generally a complex number whose phase depends on the direction of principal axis and shear deformation. Therefore, the modified DOS of BLG not only depends on the magnitude of $w$, it is very sensitive to the phase of complex parameter as well.

 Fig. S3 (b) shows the bandstructure near the valance band edge for $D = 0.4$ V/nm with $w = \pm 6$ meV and $w=\pm 6i$ meV. 
Clearly, the type of distortion in bandstructure is dependent on $w$. This is also reflected in the DOS (Fig.~S3 (c)) where there are two van-Hove singularities for $w = \pm 6i$ meV and $w = 6$ meV whereas there is only one LT for $w = -6$ meV.
Fig.~S3 (d) shows the fitting of Mott formula for $D = 0.4$ V/nm with the inclusion of different values of $w$. $w = \pm 6i$ meV and $w = 6$ meV seem to reproduce the plateau fairly well whereas $w = -6$ meV does not replicate the plateau at the same number density.
\begin{figure}
\includegraphics[width=0.9\linewidth]{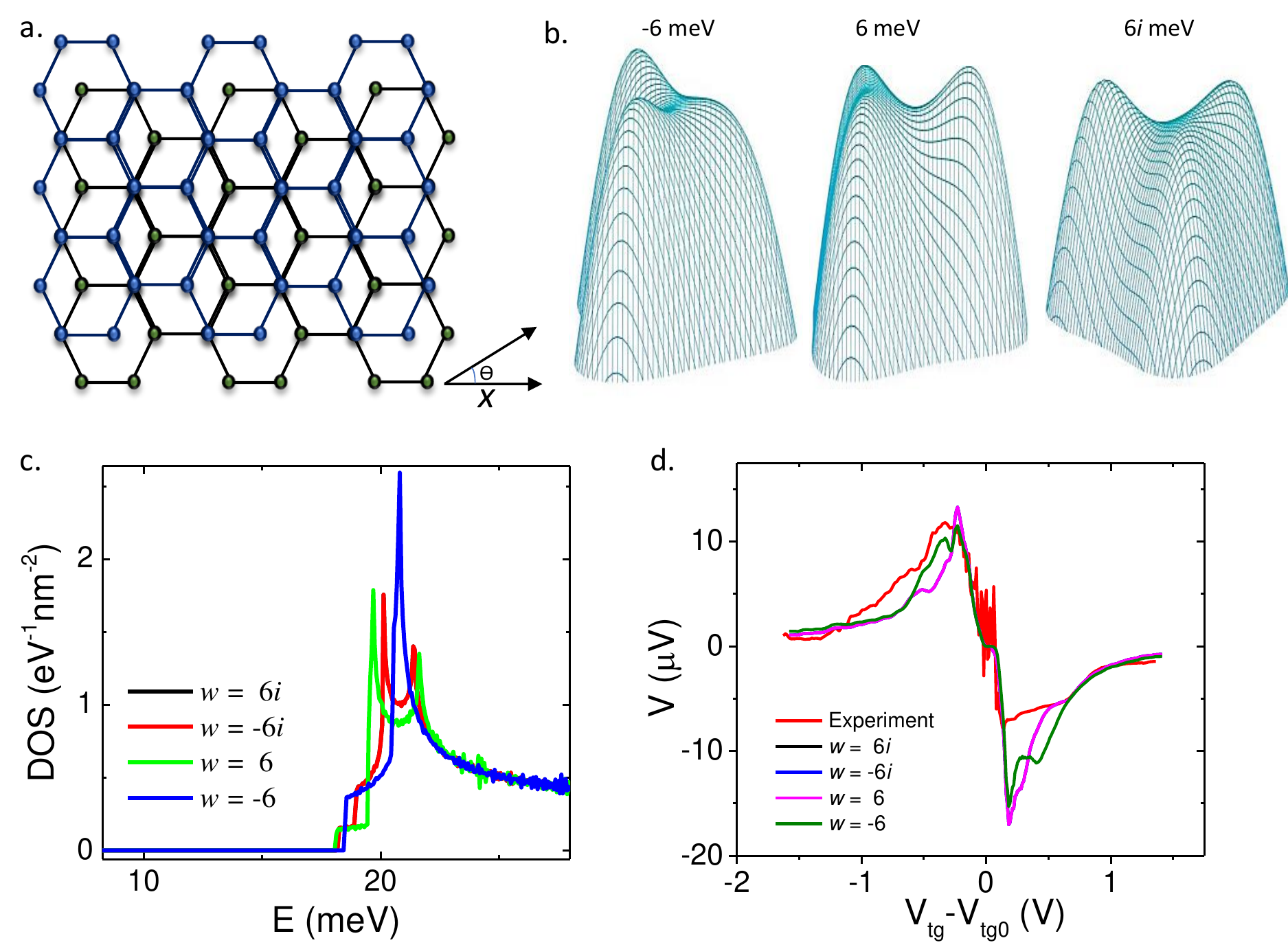}
\caption{\textbf{a.} Schematic of Bernal stacked BLG showing the angle between the coordinate axis and principal strain axis ($\theta$). \textbf{b.} Bandstructure of BLG ($D=0.4$ V/nm) for $w=6i$ meV and $w=\pm6$ meV.  \textbf{c.} DOS of BLG for various values of $w$. \textbf{d.} Fitting of Mott formula for $D=0.4$ V/nm for different values of $w$.}
\end{figure}
\label{FigS3}

\subsection{\large{\textbf{4.} Mott fitting of TEP for $D=0.2$ V/nm and $D=0.5$ V/nm}}

The analysis and fitting of Mott formula for $D=0.2$ V/nm and 0.5 V/nm is depicted in Fig.~S4. The magnified plots (Fig.~S4 (c)) show excellent reproduction of plateau on the electron doped region. The plateaus on the hole doped regime, however, are not replicated at the same density. The fitting of Mott formula with a higher value of $w$ (= 12 meV) seems to fit the plateau on the hole side perfectly (Fig.~S4 (d) for $D = 0.2$ V/nm). This electron-hole asymmetry in the value of $n_{\mathrm{L}}$ could be because of strain inhomogeneities or the presence of perpendicular strain associated with the change in interlayer distance, which has not been incorporated in our calculations~\cite{verberck2012strain}. 
\begin{figure}[H]
\includegraphics[width=1\linewidth]{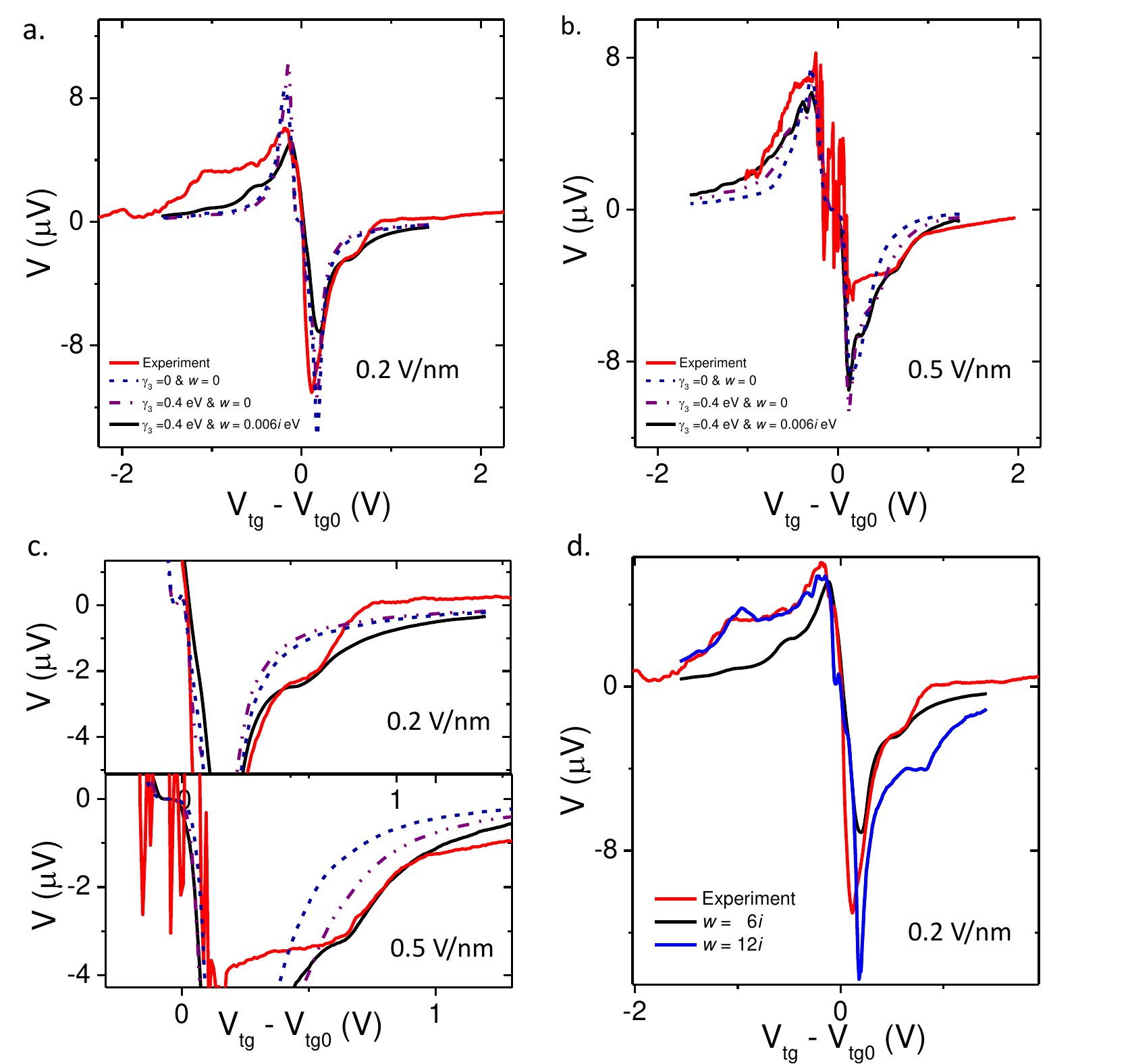}
\caption{Mott fitting of TEP with various approximations of DOS for \textbf{a.} $D = 0.2$ V/nm and \textbf{b.} $D = 0.5$ V/nm. \textbf{c.} Magnified version showing the fits for $n > 0$. \textbf{d.} Mott fits for $D = 0.2$ V/nm with $w = 6i$ meV and $w = 12i$ meV.}
\end{figure}
\label{FigS4a}